\documentclass[showpacs,twocolumn,preprintnumbers,amsmath,amssymb,pra]{revtex4}

\usepackage{graphicx}% Include figure files
\usepackage{dcolumn}% Align table columns on decimal point
\usepackage{bm}% bold math
\usepackage{color}
\usepackage{ulem}
\begin{document}

\title{Threshold non-linear absorption in Zeeman transitions.}

\author{Andal Narayanan$^{1}$}
\email{andal@rri.res.in}
\author{Abheera Hazra$^{1}$}
\author{S. N. Sandhya$^{2}$}

\affiliation{%
$^{1}$Raman Research Institute,  Sadashivnagar, Bangalore, INDIA 560 080 \\
$^{2}$Department of Physics, IIT Kanpur, Kanpur, INDIA 208 016 }

\date{\today}

\begin{abstract}
We experimentally study the absorption spectroscopy
from a collection of gaseous  $^{87}$ Rb
atoms at room temperature irradiated with
three fields.
Two of these fields are in a pump probe saturation
absorption configuration. The third field co-propagates with the pump
field.
The three fields address
Zeeman degenerate transitions between hyperfine levels
$5S_{1/2}, F = 1$ and $5P_{3/2}, F = 0$,$F =1$ around the D2 line.
We find a sub-natural absorption resonance in the counter-propagating probe
field for equal detunings of all the three fields.
The novel feature of this absorption is its abrupt onset in the vicinity of
$5P_{3/2}, F = 0$,
as the laser frequency is scanned from $5P_{3/2}, F = 0$ to $5P_{3/2}, F =1$.
The experimental results are compared with the theory modeled
after a four level system. There is a qualitative agreement between our theory
and experiment. We find the threshold absorption to be a result of
the off-resonant interaction of the strong field
with nearby hyperfine levels.
\end{abstract}


\maketitle

\section {Introduction}
The absorption spectrum of a driven degenerate two level system (DTLS)
is shown \cite{akulshin1} to be radically different from the usual Mollow triplet
corresponding to a pure two
level system. The atomic coherence induced by multi-photon interactions
in DTLS results in a competition between
constructive or destructive quantum interference
which gives rise to subnatural
width resonances (SNWR) \cite{akulshin2}. The
resonance could manifest as induced absorption (EIA), transperency (EIT),
or even gain, depending
on the polarization, intensity and detunings of the driving fields. The line
shapes are further influenced on whether the system is closed (cyclic) or
open \cite{Renzoni}, velocity selection \cite{velocity}, 
and the presence of small polarization admixtures \cite{wasik}.
Akulshin et al.,
\cite{akulshin3} have extensively studied DTLS 
involving various ground state and
excited state hyperfine levels both theoretically and experimentally.
They arrive at a criteria \cite{akulshin4} for the occurence of EIA in DTLS, 
namely (i) F$_e$ = F$_g$ + 1, 
(ii) the ground to excited state transition is closed and
(iii) F$_g > 0 $.\\
\indent The occurence of EIA has been attributed to the transfer of
coherence from excited state to the ground state \cite{Taichenechev} and also
to the transfer of population \cite{wilson}. Inhibition of EIA due to
excited state
decoherence has also been reported experimentally \cite{lezama}. Most of these
studies involve the interaction of DTLS with two driving fields. Multiple
driving fields interacting with DTLS has been reported in the context of
coherent hole burning \cite{chb} 
 where the same transition is adressed by a saturating
beam and a probe.\\
\indent In this paper we experimentally study the occurence of sub-natural
absorption resonances in the counter probe during a three field irradiation
of gaseous, room temperature $^{87} Rb.$ atoms. These three fields address
Zeeman degenerate transitions in a DTLS configuration.
These resonances arise as a result of dark resonances 
created by the co-propagating fields, inducing a higher order,
off-resonant absorption to neighbouring hyperfine levels in the counter field. 
Since the intensity regime of the probe in which the absorption is seen
is well below saturation,
we do not expect coherent population oscillations \cite{cpo} to affect
the absorption line shape \cite{velocity}. 
The absorption is studied as a function of
detunings of the three driving fields. We find, as a function of this detuning,
the existence of a sharp thresold
where the absorption begins to show up. 
We compare our experimental results with
a theoretical model comprising of a modified 'N' system which has contributions
from a double lamda system and double 'V' system depending on the detunings of the laser beams. The modeling
reproduces the threshold behaviour very clearly.
\section {Experiment}
\subsection{Level Structure}
The hyperfine levels around the D2 line of $^{87} Rb.$ is shown in Figure 1(a).
Figure 1(b) shows the Zeeman sub-levels of F = 1 ground hyperfine state and 
that of excited hyperfine states F = 0$^{\prime}$, 1$^{\prime}$ 
\footnote{Excited hyperfine states are shown primed}
relevant for the experiment described here.
\begin{figure}
\includegraphics[scale = 0.4]{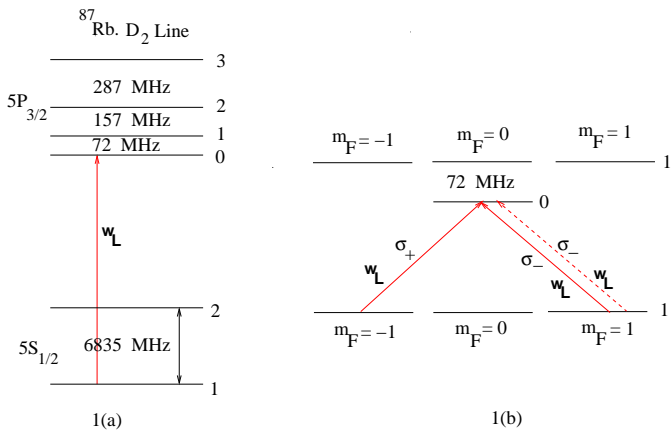}% Here is how to import EPS art
\caption{\label{fig:epsfig1} Level Scheme.}
\end{figure}
\subsection{Experimental setup}
Three beams L1, L2 and L3, derived from a single laser irradiate a 
room temperature sample
of $^{87} Rb.$ atoms. The laser is an External Cavity Diode Laser
(ECDL) with a line width typically around 1 MHz operating
around the D2 line  at 780 nm. The laser frequency is scanned around the 
hyperfine manifold starting from ground state 
$5S_{1/2}, F = 1$ to excited states
$5P_{3/2}, F = 0^{\prime}$ and $5P_{3/2}, F = 1^{\prime}$. The laser
frequency is monitored through a saturation absorption setup
as indicated in Figure 2. A phase sensitive feed back mechanism
\begin{figure}
\includegraphics[scale = 0.4]{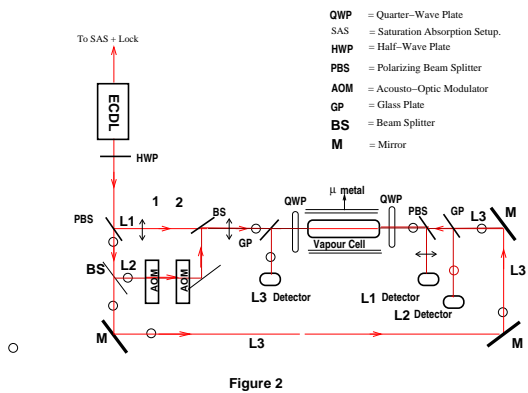}% Here is how to import EPS art
\caption{\label{fig:epsfig2}
Schematic of the experimental setup.}
\end{figure}
is employed to lock the frequency of the laser to any one 
of the excited states. 
As shown in Figure 2, L1 and L2 are collinear 
and co-propagating and L3
counter-propagates to both L1 and L2.  L1 is typically of higher 
intensity than L2 with 
intensity ranging between 1 $I_{s}$ and 5 $I_{s}$ 
where $I_s$ is the saturation intensity for the
transition $5S_{1/2}, F = 1$ and $5P_{3/2}, F = 0^{\prime}$. 
L2 has intensity in the range of 0.1 $I_s$ to
0.5 $I_s$. L3 has the least intensity ranging between 
0.1 $I_s$ and 0.01 $I_s$. All 
the three beams being derived from a single laser
have the same laser frequency $\omega_L$. At any given $\omega_L$, the  
L2 beam can be ramped over a frequency range of $\pm 10$ MHz 
around $\omega_{L}$,  by 
a combination of two Acousto-Optic-Modulators (AOMs in Figure 2). 
Thus for a given detuning $\delta_{10}$, measured from 
$5P_{3/2}, F = 0^{\prime}$ 
level, the L2 beam, through the AOMs can
scan a width of $\pm 10 MHz$ about $\delta_{10}$.
The experiment consists of  recording the transmitted intensities
of all three beams using fast photodiodes.
The transmitted intensities show two kind of variations.One of these is 
due to changing $\omega_L$. These variations mimic the  standard
saturation absorption variations in the transmission of 
counter-propagating L3 probe, for appropriate
intensity ratios of L1 and L3. But because of the presence of L2 
co-propagating
with L1, at every $\omega_{L}$, the condition for a two-photon
Raman resonance is satisfied
between L1 and L2. The results in transmission changes 
in L2 and L3 which are very different from a standard saturation
experiment.In fact, during
the $\pm 10 MHz$ frequency scan of L2 centered at $\omega_{L}$, 
we see enhanced transmission of L2 due to EIT as expected.
As the AOM scan rate is faster than the scan of laser frequency  $\omega_{L}$,
the spectral features are dominated by features during the faster 
frequency scan in the narrow bandwidth centered around
$\delta_{10}$. The slower scan of the laser frequency $\omega_{L}$ results in a 
varying background.\\
The L1 and L2 beams are orthogonally polarized with the quantization axis 
as the direction of propagation
and the L3 beam is polarized similar to the L1 
beam. The vapor cell containing gaseous Rubidium atoms in their
natural isotopic abundance, is covered with three layers of $\mu$ metal shield.
The residual field is in the range of 10 to 20 milligauss inside the cell.
\subsection{Level structure for multi-photon transitions}
The three laser beams L1, L2 and L3 can drive Zeeman degenerate transitions 
as shown in Figure 3. Figure 3(a) shows transitions 
between the Zeeman sublevels of
$5S_{1/2}, F = 1$ and $5P_{3/2}, F = 0^{\prime}$. 
The laser beams L1 (shown thick) and L2 which are 
oppositely circularly polarized form a $\Lambda$ system with the 
magnetic hyperfine levels $m_F$ = $\pm 1$ of ground state $F = 1$ and the 
excited state level F = $0^{\prime}$.
The beam L3 is counter propagating to L1 and L2  and hence there is a frequency 
offset in the atomic rest frame due to the thermal motion.
L3 connects the same set of 
levels as L1, shown here in dotted lines.
Figure 3(b) show transitions between the Zeeman sublevels of $5S_{1/2}, F = 1$ 
and $5P_{3/2}, F = 1^{\prime}$. For oppositely circularly polarized L1 and L2
this gives rise to a $\Lambda$ and a 'V' configuration for EIT.
Even here because the L3 beam (shown in dotted lines) 
is counter to L1 and L2, it does not take part in the transparency effect. \\

\indent Because of the presence of several velocity classes of atoms 
at room temperature,
even when the laser is addressing resonantly the 
transition shown in Figure 3(a), it is
off-resonantly addressing transitions in the 
F = $1^{\prime}$ manifold and vice-versa. 
This effect is most prominent for the L1 laser which has a higher intensity
than L2 and L3.
So a realistic picture of possible laser transitions is shown in Figure 3 (c). 
Here the dot-dash line indicates the transitions addressed by L1 laser
off-resonantly.
\begin{figure}
\includegraphics[scale = 0.35]{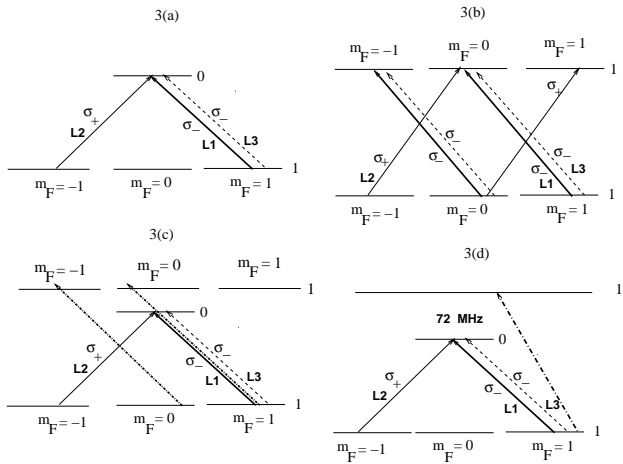}% Here is how to import EPS art
\caption{\label{fig:epsfig3}
Thick line denotes L1 laser. Dotted line denotes L3 laser. Dot-Dash line
denotes off-resonant transitions of L1 laser. L2 is denoted by an ordinary
line.
(a) Zeeman transitions between F = 1 and F = 0$^{\prime}$. (b) Zeeman 
transitions between F = 1 and F = 1$^{\prime}$. (c) Off-resonant transitions
of the L1 laser when it is addressing F = 1 to F = 0$^{\prime}$ transition.
(d) N system with resonant and off-resonant transitions.}
\end{figure}
\section {Results and Discussion} 
Figure 4 shows the typical transmitted intensities of 
L1 and L3 beams for various detunings
$\delta_{10}$. As stated before, L1, L2 and L3 beams have the 
same detuning $\delta_{10}$ as they are derived from the same laser. 
 The L2 beam scans, at every detuning 
$\delta_{10}$, a frequency range of $\pm 10 MHz$ centered
around $\delta_{10}$. 
Trace B of Figure 4(a) shows the ramp voltage scan applied to the AOMs in
the path of L2 resulting in a frequency scan of  
a $\pm 10$ MHz of L2, centered around a given $\delta_{10}$. 
On either side of these AOM frequency scans are shown Traces A and C.
Traces A and C  show the transmitted intensity profiles of L2 and L3
respectively, 
during the scan of L2 centered around a given value of $\delta_{10}$. 
As can be seen from trace A, L2
shows increased transmission over the background, 
whenever L1 and the scanned L2 are at
the same detuning, satisfying the two photon resonance condition for EIT. 
The EIT window is typically of 1 - 2 MHz width.
On the other hand, the transmitted 
intensity of L3 given in Trace C shows a sharp absorption resonance, 
at the very position at which L2 shows EIT transmission resonances. 
The width of this absorption is sub-natural ranging
between 300 KHz to 800 KHz. 
\begin{figure}
\includegraphics[scale = 0.4]{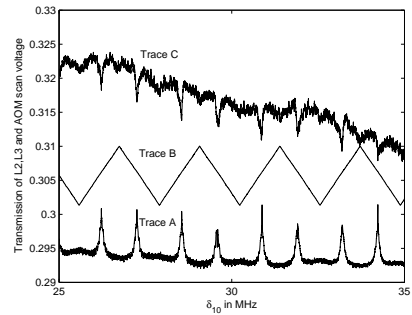}% Here is how to import EPS art
\caption{\label{fig:epsfig4}
Trace B denotes the ramp voltage going to the AOM of the L2 beam 
resulting in a frequency scan of $\pm 10 MHz$ around a given $\delta_{10}$.
Trace A denotes L2 transmission showing increased transmission
whenever the two co-propagating L1 and the scanned 
L2 meet the two-photon resonance condition for EIT. Trace C shown decreased
transmittance (absorption) of the counter L3 beam at two-photon resonance
}
\end{figure}
The graph of Trace C has been amplified 6 times for clear representation
in this combined graph.
Since this absorption in L3 occurs whenever the co-propagating
L1 and L2 satisfy the two-photon EIT condition, it implies 
 a  higher order
absorption mechanism for L3 absorption. 
In fact, a related three-photon absorption was 
seen by us in an N level scheme ~\cite{Andal1} for similar geometry of beams. 
The novelty of such a feature here, for Zeeman degenerate transitions, 
is its abrupt onset ~\cite{Andal2} as is shown in Figure 5. \\
\begin{figure}
\includegraphics[scale = 0.5]{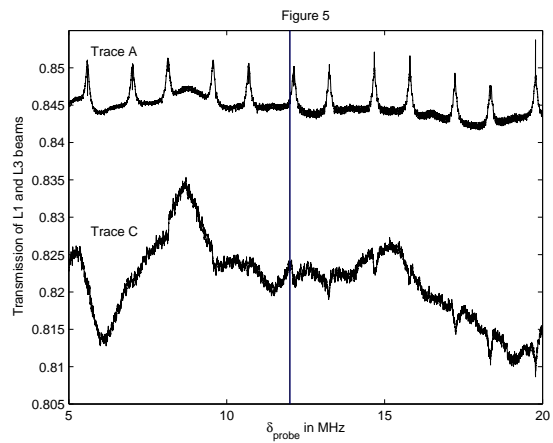}% Here is how to import EPS art
\caption{\label{fig:epsfig5} Traces A and C of Figure 4 shown in a region
of $\delta_{10}$ around 12 MHz showing abrupt onset of L3 absorption
resonance}
\end{figure}
\indent Shown in Figure 5 are Traces A and C from the same experimental run 
as it 
is for Figure 4 but in a different region of $\delta_{10}$ around 
$F = 0^{\prime}$. 
As can be seen from this graph, while Trace A remains qualitatively unchanged 
even in this region, Trace C shows abrupt dissappearence of the absorption
feature below $\delta_{10} < 12$ MHz marked by the vertical line in the figure.
For all values of $\delta_{10}$ below this value, there is no narrow absorption
seen in Trace C. The background absorption of L3 in this region show changes
which is not correlated with the appearence of EIT feature in the L2 field
(Trace A). We have estimated for this graph that the narrow 
absorption dissappers within 
a 5 MHz change of $\delta_{10}$ to the left of the vertical line. 
Conversly, we can say that the 
 nonlinear
absorption in L3 has an abrupt onset around $\delta_{10}$ = 12 MHz.\\
\indent For the sake of clarity we have shown in Figures 6(a) and 6(b), 
the entire region 
of $\delta_{10}$ values extending upto -35 MHz. The narrow 
absorption feature in L3 (Trace C) is totally absent in this region. 
Only beyond the vertical line do we see the onset of this absorption.
On the contrary Trace A,of this figure, representing L2 transmission,
continues to show the EIT feature for all values of $\delta_{10}$.
Figure 6(b) shows that the narrow absorption feature in Trace C
continuing for positive $\delta_{10}$ atleast upto 35 MHz.\\
\begin{figure}
\includegraphics[scale = 0.5]{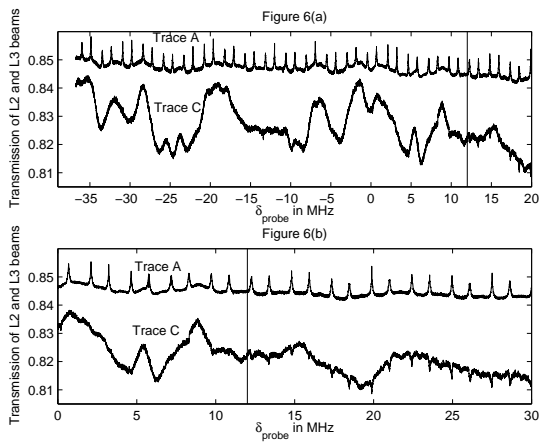}% Here is how to import EPS art
\caption{\label{fig:epsfig6}
(a) Absence of L3 absorption resonance to the left of the vertical line
(Trace C) for values
extending upto $\delta_{10}$= $- 35 MHz$. (b)  Presence of L3 absorption
resonance (Trace C) beyond the vertical line for $\delta_{10}$ = 35 MHz.
Trace A in both the figures represents L2 beam transmission showing EIT for
all values.}
\end{figure}
\indent We have repeated the experiment with various intensities of the L2 
field keeping the L3 intensity to be the same. 
Every time we see that there is an abrupt onset of EIT correlated 
absorption in L3 occuring in the vicinity of $\delta_{10}$. The value of 
$\delta_{10}$ at which this occurs is not strongly
dependent on the intensity of the L1 beam. However, the width of the absorption
feature and the EIT transmission feature increase at higher intensities
of L2 as expected.
Plotted in Figure 7 is a typical absorption contrast of the narrow
absorption in L3 as a function of $\delta_{10}$. We see in this figure that
the absorption rises to its maximum value in a narrow range of $\delta_{10}$.
\begin{figure}
\includegraphics[scale = 0.4]{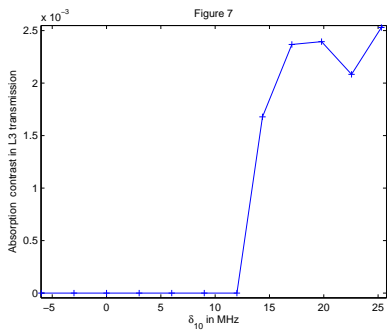}% Here is how to import EPS art
\caption{\label{fig:epsfig7}
Contrast of L3 absorption resonance showing a sharp threshold around 
$\delta_{10} = 12 MHz$.
}
\end{figure}
\subsection{Numerical Modeling}
As seen from Figure 4, for intensities of L2 and L3 beams well within $I_s$, the absorption
feature in L3 occurs only to the blue of $F = 0^{\prime}$. 
To understand this effect we tried to model our system as a four level system with two ground levels as $m_F = \pm 1$
ground state Zeeman sub-levels of $5S_{1/2}, F = 1$ and two excited levels being the hyperfine
levels $5P_{3/2}, F = 0^{\prime}$ and $5P_{3/2}, F = 1^{\prime}$. The Hamiltonian consists of 
all the various transitions as depicted in Figure 3(d). The counter-propagating weak beam is 
treated perturbatively retaining terms upto  first order in perturbation.
This is done explicitly to seperate the contribution to the density matrix elements from the 
forward L1 and counter L3 beams which address the same set of transitions. More importantly,
off-resonant transitions are taken into account for the strong L1 beam. \\
\indent We divide the total Hamiltonian into two parts.
\begin{equation}
H = H_{0} + \Delta H
\end{equation}
Where $H_{0}$ is the Hamiltonian describing the four-level system
interacting with the L1 and L2 fields.
$\Delta H$ has L3 interaction terms only.
The Liouville equation for the total Hamiltonian has the form
\begin{equation}
\frac{d(\rho + \Delta \rho)}{dt} = -\frac{i}{\hbar} [H_{0} + \Delta H,\rho + 
\Delta \rho] -\frac{1}{2} \{\Gamma, \rho + \Delta \rho\}  \label{lioi1}
\end{equation}
Neglecting terms which are second order in $\Delta \rho$ and $\Delta H$, the abo
ve equation becomes,
\begin{equation}
\frac{d(\rho + \Delta \rho)}{dt} = -\frac{i}{\hbar} [H_0, \rho] -\frac{i}{\hbar}
 [H_0, \Delta \rho] -\frac{i}{\hbar} [\Delta H, \rho] -\frac{1}{2} \{\Gamma, 
\rho + \Delta \rho\}  \label{lio2}
\end{equation}
We know that for the unperturbed system the equation
\begin{equation}
\frac{d(\rho)}{dt} = -\frac{i}{\hbar} [H_0, \rho] -\frac{1}{2} \{\Gamma, \rho\}
\label{lio3}
\end{equation}
is satisfied. We proceed to find numerical solution for steady state values of
$\rho$ solving fifteen coupled equations with the constraint
\begin{equation}
\sum_{i} \rho_{ii} = 1
\end{equation}
 We now solve the equation for $\Delta \rho$
\begin{equation}
\frac{d(\Delta \rho)}{dt} = -\frac{i}{\hbar} [H_0, \Delta \rho] -\frac{i}{\hbar}
 [\Delta H, \rho] -\frac{1}{2} \{\Gamma, \Delta \rho\}  \label{lio4}
\end{equation}
where the $\rho$ values are given as inputs from the solution obtained by solvin
g (4).
Imposing the constraint,
\begin{equation}
\sum_{i} \Delta \rho_{ii} = 0
\end{equation}
we get 15 coupled equations for $\Delta \rho$ which are numerically solved to 
obtain steady state density matrix values for the weak counter-propagating 
L3 field.\\
\indent Figure 8 shows the result of our calculations. In Figure 8(a) we give
 the normalised transmission of the co-propagating scanned L2 beam. 
As expected, for all detunings $\delta_{10}$ this 
shows transparency as is the case with our experimental result shown in Trace A
of Figure 5. The transmission of the  L3 beam is shown in 8(b). This 
shows gain for values of $\delta_{10} <= 0$, and absorption for values of 
$\delta_{10} > 0$. This should be compared with the experimental curve for L3
transmission (Trace C) of Figure 5. While we do see absorption for values
of $\delta_{10} > 0$, for values of $\delta_{10} < 0$, 
we do not see the theoretically
predicted gain (Figure 8(b)) in our experimental curve (Trace C of Figure 5).
We think that this could be due to the presence of a non-vanishing $\pi$
component in our counter L3 beam.
\begin{figure}
\includegraphics[scale = 0.5]{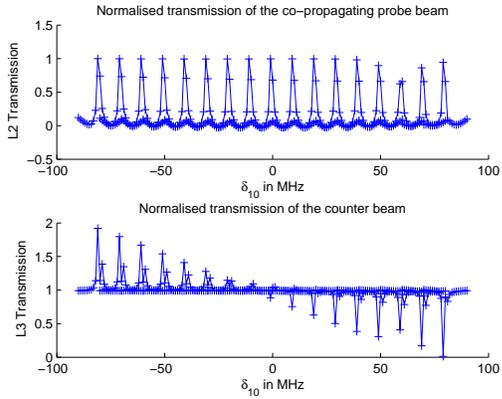}% Here is how to import EPS art
\caption{\label{fig:epsfig8}
Transmission of L2 and L3 for experimental
 parameters values as depicted in Figure 5.
(a) Normalized transmission of L2 beam. (b) Normalized transmission of
L3 beam
}
\end{figure}
\subsection{Threshold non-linear resonances with off-resonant interactions}
It is well known that for a pure $\Lambda$ or a double $\Lambda$ system, in the absence of 
any perturbing mechanism introducing light shifts to a dark state, 
the system exhibits no higher order non-linearity 
~\cite{phasegate}.
In our system, this perturbation is introduced by the off-resonant transitions 
addressed
by the strong L1 laser. This laser makes the transition probability 
for transitions of suitable velocity class of atoms,
to the F = 1$^{\prime}$, non-negiligible. 
Our system, due to this off-resonant interaction, can be seen
as modified N system as shown in
Figure 3(d). So, for a suitable velocity class of atoms,
there is a suppression of linear
susceptibility due to EIT and a simultaneous enhancement of absorption 
due to higher order 
nonlinearity to the nearby hyperfine level. This absorption due to
higher order non-linearity, is distinctly seen in L3 beam due to its
counter geometry.
In effect, this feature is 
very similar to the higher order non-linear absorption seen in 
N systems with subnatural widths 
~\cite{Nreference1,Nreference2,Nreference3}.
The difference in the present case is its strong dependence
on the detuning. \\ 
\indent It is well known that the DTLS systems show gain  ~\cite{akulshin1} in some intensity regime.
According to the theoretical model, since our laser addresses a  DTLS 
there is gain due to the contribution from the double lambda part.
In addition, there is also nonlinear absorption due to off-resonant interaction which is dominant.
The gain becomes negligible as the laser goes beyond $\delta_{10} = 0^{\prime}$.
\color{black}(Figure 8(b)). This results in absorption of L3 beam to be seen 
prominently, only for values of $\delta_{01} >= 0^{\prime}$.\\
\indent We realised from our theoretical simulations that the main reason
for the threshold absorption is the off-resonant addressing of F=$1^{\prime}$
level by our strong L1 beam. When we make this off-resonant contribution  
to be zero, then we see that both L2 and L3 transmission show increased
transmission for all $\delta_{0}$, with similar widths. 
This is shown in Figure 9.
\begin{figure}
\includegraphics[scale = 0.5]{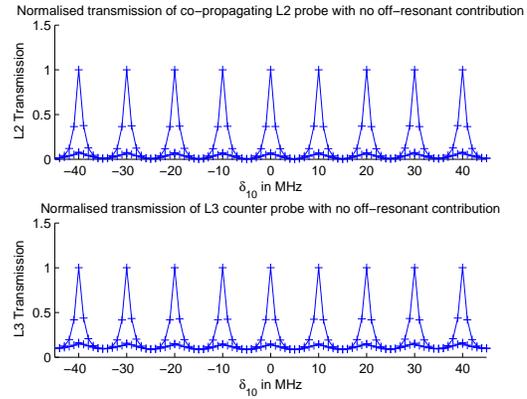}% Here is how to import EPS art
\caption{\label{fig:epsfig9}
Theoretical transmissions of L2 and L3 beams for experimental
 parameters values as depicted in Figure 5 when off-resonant contribution
due to L1 laser is made zero.
(a) Normalized transmission of L2 beam. (b) Normalized transmission of
L3 beam
}
\end{figure}

Both the double $\Lambda$ and Tripod systems have shown to have 
many properties ~\cite{Tripod1,Tripod2,Tripod3} 
which aid specific quantum engineering of 
superposition states ~\cite{engineer}. 
Our system which is  a modified double $\Lambda$ system 
exhibits a threshold non-linear absorption of sub-natural width at the
same frequency where transparency is seen due to ground state Zeeman
coherence. The seperation of absorptive and transmittive features in
L2 and L3 beams is achieved by geometry of beam propagation.
Such a feature is desirable from the view point of 
monitoring the fidelity of the dark states. Passive monitoring 
of counter propagating probe should reveal the nature of CPT coherence 
in the forward beams. Also, by using the real part of the non-linear 
susceptibility,  one can design phase gates
which work only  in a specific bandwidth. In this sense, the threshold 
phenomenon seen here can be used as a frequency filter.
\section{Conclusions}
We report in this paper an experimental observation of a sub-natural absorption
feature in a 
modified N system around Zeeman degenerate transitions in the
D2 manifold of $^{87} Rb$. This feature is seen when the gaseous Rb. atoms are 
irradiated with three fields all derived from the same laser. 
The novelty of this absorption resonance is its
abrupt onset beyond a certain detuning of the laser. 
We show that this absorption feature in one of the beams, 
arises due to off-resonant absorption 
to the neighbouring hyperfine excited states through a higher order absorptive
non-linearity. 
 We have modeled
our system using density matrix formalism. This shows that the threshold
nature of this absorption is due to a competition between gain present in
such DTLS and off-resonant coherence induced absorption. 
This feature and its distinct appearence in only one of the three fields
provides a powerful tool for frequency filtering in wave guides.
\section{ Acknowledgements} 
One of us, SNS thanks the Department of Science and Technology, India, for financial assitance through the WOS-A scheme.
\bibliographystyle{apsrev}

\begin{thebibliography}{100}
                                                                                
\bibitem{akulshin1} A. Lipsich, S. Barreiro, A.M. Akulshin, and A. Lezama, 
Phys. Rev. A {\bf 61} 053803 (2000).
                                                                                
\bibitem{akulshin2} A.M. Akulshin, S. Barreiro and A. Lezama,
 Phys. Rev. A {\bf 57} 2996 (1998)
                                                                                
\bibitem{akulshin3} A. Lezama , S. Barreiro, A. Lipsich, and A. M. Akulshin,  Phys. Rev. A {\bf 61} 013801 (1999).
                                                                                
\bibitem{akulshin4} A. Lezama, S. Barreiro, and A. M. Akulshin, Phys. Rev. A {\bf 59} 4732 (1999).
                                                                                
\bibitem{Renzoni} F. Renzoni, W. Maichen, L. Windholz, and E. Arimondo, Phys. Rev. A {\bf 55} 3710 (1997).
                                                                                
\bibitem{wasik} G Wasik, W Gawlik, J Zachorowski and Z Kowal, Phys. Rev. A {\bf 64} 051802(R) (2000).
                                                                                
\bibitem{velocity}  S. Haroche and F. Hartmann, Phys. Rev. A {\bf 6} 1280 (1972);
C. Affolderbach, S. Knappe, R. Wynands, A. V. Taicenachev
and V. I. Yudin, Phys. Rev. A {\bf 65 } 043810 (2002).
                                                                                
\bibitem{Taichenechev} A. V. Taichenachev, A. M. Tumaikin, and V. I. Yudin, Phys. Rev. A {\bf 61} 011802 (R) (1999).
                                                                                
\bibitem{wilson} C. Goren, A. D. Wilson-Gordon, M. Rosenbluh and H. Freidmann,
Phys. Rev. A {\bf 72} 023826 (2005); {\it ibid}, {\bf 70} 043841 (2004);
{\it ibid}, {\bf 69} 053818 (2004); {\it ibid} {\bf 67 } 033807 (2003).
                                                                                
\bibitem{lezama} G. Ban, V. Lorent and A. Lezama, Phys. Rev. A. {\bf 67}
043810 (2003).
                                                                                
\bibitem{chb} Y. Gu, Q. Sun, Q. Gong, Phys. Rev. A {\bf 69} 063805 (2004).

\bibitem{cpo} R. W. Boyd, M. G. Raymer, P. Narum and D. J. Harter, 
Phys. Rev. A {\bf 24} 411 (1981).
\bibitem{Andal1}A. Narayanan, A. Sharma, Preethi T.M, A. Hazra and H. Ramachandran, accepted for publication in Canadian Journal of Physics.

\bibitem{Andal2}A. Hazra, A. Narayanan and S.N. Sandhya, Poster presentation
at {\it ICAP 2008}, Storrs, Connecticut (2008).

\bibitem{phasegate} S. Rebic , D. Vitali, C. Ottaviani, P. Tombesi, M. Artoni, F. Cataliotti, and R. Corbalan,  Phys. Rev. A.,  {\bf 70}, 032317 (2004)

\bibitem{Nreference1} S. Zibrov, 
I. Novikova, D. F. Phillips, A. V. Taichenachev,
V. I. Yudin, R. L. Walsworth and A. S. Zibrov, 
Phys. Rev. A, {\bf 72}, 011801(R) (2005)

\bibitem{Nreference2} A.S. Zibrov, C.Y. Ye, Y. V. Rostovtsev, A. B. Matsko, and M. O. Scully,  Phys. Rev. A, {\bf 65}, 043817 (2002)

\bibitem{Nreference3} C. Y. Ye, A. S. Zibrov1, Yu. V. Rostovtsev, and M. O. Scully,  Phys. Rev. A, {\bf 65}, 043805 (2002).

\bibitem{Tripod1} D. Petrosyan and Y P.Malakyan, Phys. Rev. A, {\bf 70}, 023822, 
(2004)

\bibitem{Tripod2}A. M. Akulshin, A I Sidorov, R.J McLean and P Hannaford, J. Opt.B, {\bf 6} 491 (2004)
                                                                                
\bibitem{Tripod3} Y Han, J Xiao, Y Liu, C Zhang, H Wang, Min Xiao, and K Peng, Phys. Rev. A, {\bf 77}, 023824 (2008)

\bibitem{engineer} F Vewinger , M Heinz, R G Fernandez, N V. Vitanov , and K Bergmann, Phys. Rev. Lett., {\bf 91}, 213001 (2003).
\end{thebibliography}

\end{document}